\documentclass{article}[12pt]

\usepackage{natbib}
\usepackage{longtable}
\usepackage{url}
\usepackage{graphics}
\usepackage{graphicx}
\usepackage{amsmath}
\usepackage{geometry}
\usepackage{epstopdf}

\begin{document}

\title{Descriptive Statistics of the Genome: \\Phylogenetic Classification of Viruses} 
\author{Troy Hernandez$^*$\\
Data Science\\
Sears Holdings Company\\
1 N. State St.\\
Chicago IL, 60602\\
\texttt{troy.hernandez.phd@gmail.com}\\
\bigskip
\and
Jie Yang\\%
Department of Mathematics, Statistics, and Computer Science\\
University of Illinois at Chicago\\
322 Science and Engineering Offices\\
851 S. Morgan Street, Chicago, IL 60607-7045\\
\texttt{jyang06@uic.edu}
}

\date{}



\maketitle

\bigskip

\bigskip

\begin{center}
Keywords: alignment-free, phylogenetics, virology, machine learning, classification
\end{center}

\newpage

\begin{abstract}

The typical process for classifying and submitting a newly sequenced virus to the NCBI database involves two steps.  First, a BLAST search is performed to determine likely family candidates.  That is followed by checking the candidate families with the Pairwise Sequence Alignment tool for similar species.  The submitter's judgement is then used to determine the most likely species classification.  The aim of this paper is to show that this process can be automated into a fast, accurate, one-step process using the proposed alignment-free method and properly implemented machine learning techniques.


We present a new family of alignment-free vectorizations of the genome, the generalized vector, that maintains the speed of existing alignment-free methods while outperforming all available methods.  This new alignment-free vectorization uses the frequency of genomic words (k-mers), as is done in the composition vector, and incorporates descriptive statistics of those k-mers' positional information, as inspired by the natural vector.

We analyze 5 different characterizations of genome similarity using $k$-nearest neighbor classification, and evaluate these on two collections of viruses totaling over 10,000 viruses.  We show that our proposed method performs better than, or as well as, other methods at every level of the phylogenetic hierarchy.

The data and R code is available upon request.

\end{abstract}

\newpage

\section{Introduction}
\label{s:intro}

\begin{quote}
At the end of the day, some machine learning projects succeed and some fail. What makes the difference? Easily the most important factor is the features used.\\
-Paul Domingos~\protect{\citep{domingos}}
\end{quote}

The proliferation of low-cost, high-speed genomic sequencing technology has and will continue to give the scientific community ever-increasing amounts of genomic data.  Experts will no longer have the ability to manually classify this torrent of biological data.  Automated virus classification systems have begun appearing in the past few years to assist experts and practitioners \citep{pascweb,gail,webnv}.  These classification systems rely broadly on two different measures of similarity between the genome; sequence alignment identity and alignment-free vectorizations.

Virus classification by pairwise sequence comparison \citep{pasc} relies on the sequence alignment identity between every pair of viruses.  For reasons of computational complexity, all pairs of viruses are aligned instead of all viruses being aligned at once as in multiple sequence aligment (MSA).  MSA (i.e. aligning entire groups of genomic sequences at once) has a computational complexity of $O(n^m)$, where $n$ is the length of a viral sequence and $m$ is the number of viruses being compared.  For this reason, all of the pairwise identities of viruses in a given family are pre-computed. After the pairwise identities of a new virus are calculated a histogram of the identity scores are displayed, color-coded according to their sub-family, genus, and species.  From there, experts use their best judgement to determine the proper subfamily, genus, and species classifications.

In the alignment-free/vectorization approach, statistics of each genome are compiled, stored, and new viruses are then classified according to various learning algorithms.  The bulk of the literature in alignment-free methods relies on the bag-of-words model, also known as k-mers within the bioinformatics community.  k-mers are genomic words from the alphabet $\{ A, C, G, T \}$ of length k; e.g. for k=3, ``AGC'', ``CTA'', and ``TAG''.  For a given k, a vector of k-mer frequencies can be used in learning algorithms for clustering or classification~\citep{vinga}.

Another alignment-free approach is the natural vector~\citep{modeng}.  The natural vector characterizes the distribution of a genome's nucleotides.  That characterization consists of the counts of $A, C, G,$ and $T$ in addition to positional information.  That is, the mean position of the nucleotides and their central moments; i.e. the $2^{nd}$, $3^{rd}$ , $4^{th}$, etc. central moments.

In this paper we extend the idea of incorporating information about the distribution of k-mer positions to a genomic vectorization.
The primary contributions are as follows:

\begin{itemize}
\item Characterizing k-mer positional distribution information in a vector via the proposed generalized vector (GV).
\item Analysis of 5 different characterizations of genome similarity; the composition vector (CV), the complete composition vector (CCV), the natural vector (NV), pairwise sequence alignment (PASC), and GV.
\item Comparative evaluation of the two collections of viruses families/genera mentioned above totaling over 10,000 unique viruses.
\end{itemize}

In section \ref{s:analysis} we describe the source, curation, and details of the data in addition to the algorithm, the implementation details, and the expectations for performance on each method.  We evaluate the different methods in section \ref{s:eval} and conclude in section \ref{sg0:concl}.

\newpage

\section{Methods}

\subsection{Related Work}
\label{s:gnvbackground}

In this section we describe various methods used in the literature to quantify similarity in genomes.  Three of the methods are alignment-free; i.e. they use statistics collected from a genome as components in a vector.  Those vectors are then used in learning algorithms for clustering or classification.  Alternatively, MSA and PASC align genomes and measure similarity directly from those alignments.  In section \ref{s:analysis}, the algorithms and pre-processing used to implement the classifications are described.  This will affect the measures of similarity differently for the different representations.

\subsubsection{Sequence Alignment}

A review of sequence alignment is beyond the scope of this paper, but one can be found in \cite{alignment}.  What is important, with regards to this paper, is the computational complexity of MSA.  Given a collection of $m$ sequences of length $n$ the complexity is $O(n^m)$.  Newer implementations have brought speed-ups beyond the naive implementation but large-scale comparisons can still be prohibitive.  PASC gets around this by aligning every pair of sequences and uses those pairwise scores for a similarity matrix.

\subsubsection{K-mers}

The bag-of-words model is ubiquitous in natural language processing~\citep{bagofwords}. In this model a text document is converted into a vector where each component represents a word.  This conversion results in the loss of grammar and word order information.  

Within bioinformatics, the bag-of-words model has been adapted to work on genomes.  The `words' in this case are sub-strings of nucleotides in the genome.  Sub-strings of length $k$, known as {\it k-mers}, can be of length 1 to $n$ for a given sequence of length $n$.  These k-mers are extracted from the sequence by sliding a window of length $k$ over the genome from the $1^{st}$ position to the $(n-k+1)^{st}$ position.  For example, in the string $S$=GATTACA 
%
there are 6 non-zero 2-mers:

$n_{AC}=1, n_{AT}=1, n_{CA}=1, n_{GA}=1, n_{TA}=1, n_{TT}=1$

This results in a vector of counts:

\begin{align}
 n_2 &= (n_{AA},n_{AC},n_{AG},n_{AT}, n_{CA},n_{CC},n_{CG},n_{CT}, \nonumber \\
 &\qquad {} n_{GA},n_{GC},n_{GG},n_{GT}, n_{TA},n_{TC},n_{TG},n_{TT}) \\
 &= (0,1,0,1,1,0,0,0,1,0,0,0,1,0,0,1)
\end{align}

Typically, by dividing by $l-k+1$ these k-mer counts are converted to frequency vectors, $f_k$.  Due to the 4 letter nucleotide alphabet, for a given $k$, there are $4^k$ components in the k-mer frequency vector.  For example:

\begin{align}
 f_2 &= (0,\dfrac{1}{6},0,\dfrac{1}{6},\dfrac{1}{6},0,0,0, 
 \dfrac{1}{6},0,0,0,\dfrac{1}{6},0,0,\dfrac{1}{6})
\end{align}

\subsubsection{Composition Vector}

It has been shown that classification using k-mers can be improved by using some informed scale and location shifts of the frequency vector \citep{hao2003}.  This is known as the composition vector (CV).  There are many different proposed parameters for the scale and location shifts.  Here we focus on a Markov Model as described in \cite{chan2010}.

For a k-mer $u$, we estimate its expected frequency using its two component $k-1$ length words.  As an example, let $u=LwR$=GATTACA, where $L$=G, $w$=ATTAC, and $R$=A.  Following \cite{chan2010}, we estimate its expected frequency:

\begin{align}
 \mathcal{P}(LwR) &= \mathcal{P}(Lw) \mathcal{P}(R \mid Lw) \\
 &\approx \mathcal{P}(Lw) \mathcal{P}(R \mid w) \\
 &= \dfrac{\mathcal{P}(Lw) \mathcal{P}(wR)}{\mathcal{P}(w)}
\end{align}

To get the composition vector component for k-mer $u$, $c_u$, we use the frequency of $u$, $f_u$, and its expected frequency, $\mathcal{P}_u$:

\begin{align}
c_u = \dfrac{f_u-\mathcal{P}_u}{\sqrt{\mathcal{P}_u}}
\end{align}

For a given $k$ this results in the composition vector:

\begin{align}
c_k = (c_{u_1}, \ldots, c_{u_{4^k}}).
\end{align}
\subsubsection{Complete Composition Vector}
\label{ccv}

The complete composition vector (CCV) takes the composition vector for various values of $k$, $c_k$, and concatenates them \citep{ccv}.  This produces the CCV:

\begin{align}
v_k = (c_1, \ldots, c_k)
\end{align}

For the CV and a fixed $k$, using the values without additional transformations is sufficient.  When using the CCV with distance matrices another transformation is necessary for the following reason:  Concatenating the CVs of a genome from $k=1 \ldots 5$, 
the vector will have 4 components from $c_1$ and $4^5=1024$ components from $c_5$.  This makes the contribution of $c_1$ negligible to the distances computed.  For this reason, as shown in section \ref{sec:RCA}, we use a transformation informed by the data.

\subsubsection{Natural Vector}

k-mers and the composition vector throw out all location information for the nucleotides, the natural vector does not.  The natural vector characterization of genomes \citep{modeng,nvpaper} consists of the counts, mean positions, and central moments of the nucleotides A,C,G, and T.  For $u=$ A, C, G, T,

\begin{itemize}

\item[(1)] Let $S=(s_1,s_2, \ldots, s_n)$ be a nucleotide sequence of length $n$; i.e. $s_i \in \{A,C,G,T\}$ for $i = 1,2, \ldots, n.$

\item[(2)] Let $n_u$ denote the number of letter $u$ in $S$ and $n$ denote the length of $S$, such that $\sum_{u} n_u = n$

\item[(3)] Let $s_u[i]$ denote the position of the $i^{th}$ letter $u$, that is

\begin{align}
s_u[1] < \cdots < s_u[n_u]
\end{align}

and

\begin{align}
S[s_u[i]]=u, \mbox{ for } i=1,\ldots, n_u.
\end{align}

\item[(4)] Let the mean position of letter $u$ be

\begin{align}
\mu_u = \sum_{i=1}^{n_u} s_u[i]/n_u
\end{align}

\item[(5)] For $j=2,\ldots, n_u$, let

\begin{align}
d^j_u = \sum_{i=1}^{n_u} \frac{(s_u[i]-\mu_u)^j}{n_u^{j-1}n^{j-1}}.
\end{align}

\end{itemize}

In theory, any number of central moments can be used.  In practice, only the second central moment (i.e. $j=2$) is used resulting in a 12-dimensional vector~\citep{nvpaper}.  This results in a vector:

\begin{align}
(n_A, \mu_A, d^2_A,\> n_C, \mu_C, d^2_C, n_G, \mu_G, d^2_G,\>n_T, \mu_T, d^2_T)
\end{align}

\subsection{Proposed Vectorization}
\label{s:gnvmethodology}

Given the k-mer, composition vector, complete composition vector, and natural vector representations of the genome, we introduce the Generalized Vector (GV).  Observing that the composition vector throws out the positional information of the genome and the natural vector retains this information, but only for k-mers of length 1, it becomes clear that a large space of descriptive statistics of the genome is being ignored.  In addition to extending the natural vector definition to k-mers with values of $k$ greater than 1, we also make some adjustments.  

\subsubsection{Coordinates of Natural Vector}
\label{nvcoordinates}

Suppose $n$ is large enough. Let $s_u$ be a randomly chosen position for the nucleotide $u$.
Assume that $s_i$ follows an \textit{iid} discrete distribution with 4 outcomes for $i=\{1,\ldots, n\}$ with proportions $(p_A, p_C, p_G, p_T)$, where $0<p_u<1, u = A, C, G, T,$ and $\sum_u p_u =1$.  Then approximately,

\begin{equation}
(s_u - \mu_u)/n \stackrel{\cdot}{\sim} {\rm Unif}(-1/2, 1/2)
\end{equation}

\begin{equation}\label{muk}
\mu_u \stackrel{\cdot}{\sim} \frac{n}{2}
\end{equation}

and

\begin{equation}\label{dkj}
 d^j_u \stackrel{\cdot}{\sim}
  \begin{cases}
   \frac{n}{2^j(j+1)n_u^{j-2}} & \text{if } j=2d \\
   0       & \text{if } j=2d-1
  \end{cases}
\end{equation}

because

\begin{align}
  \frac{1}{n_u}\sum_{i=1}^{n_u}\frac{(s_u[i]-\mu_u)^j}{n^j} &\stackrel{\cdot}{\sim} \int_{-1/2}^{1/2} x^j dx \\
  &=
    \begin{cases}
      \frac{1}{2^j(j+1)} & \text{if } j=2d \\
      0       & \text{if } j=2d-1
    \end{cases}
\end{align}

Due to the term ``$n_u^{j-2}$'' in (\ref{dkj}), which is roughly $(n p_u)^{j-2}$, $d^j_u$ will be much smaller than $n_u$ and $\mu_u$ for large $n$ and $j>2$. Therefore, the coordinates after the first 12 of the natural vector will be negligible when calculating the distances used to measure similarity.

\subsubsection{Generalized Vector}

In extending the natural vector to values of $k$ greater than 1, we first replace counts of k-mers, $n_u$, with their respective CVs, $c_u$.  The insight of the CV, which is especially important for the CCV, is that the frequencies of k-mers and (k-1)-mers are generally highly correlated \citep{ccv}.  Additionally, we concatenate the collection of CVs, $c_k$, resulting in $v_k$ as defined in section \ref{ccv}.

Secondly, we add in the length $n$.  When trying to distinguish between different families of viruses, instead of just distinguishing between different species, the length of a genome is one of the most important factors.  

Third, we use the standardized moments, $\dfrac{\mu^j}{\sigma^j}$, where $\mu^j$ represents the $j^{th}$ moment about the mean and $\sigma$ represents the standard deviation,

\begin{itemize}

\item[]

\begin{equation}
\mu_u^j = \frac{1}{n_u} \sum_{i=1}^{n_u}(s_u[i]-\mu_u)^j
\end{equation}

\item[]

\begin{equation}
\sigma_u = \sqrt{\frac{1}{n_u} \sum_{i=1}^{n_u}(s_u[i]-\mu_u)^2}
\end{equation}

\end{itemize}

This is used instead of the scaled central moments that are used in the natural vector.  In particular, $j=3$ is skewness and $j=4$ is kurtosis.  The reason for this is that
the scaling of the central moment by $\frac{1}{n^{j-1}}$ makes it so that the higher order moments converge very quickly to 0.
Lastly, similarly to CCV, we concatenate the vectors described above for various values of $k$; e.g. $k=1 \ldots 5$.
The \textit{generalized vector}, $g_k^j$, of a DNA sequence $S$ is defined by

\begin{equation}
(n, v_k, \mu_1, \ldots, \mu_k, \sigma^2_1,\ldots,\sigma^2_k,
\frac{\mu^3_1}{\sigma^3_1},\ldots,\frac{\mu^3_k}{\sigma^3_k},
\ldots,
\frac{\mu^j_1}{\sigma^j_1},\ldots,\frac{\mu^j_k}{\sigma^j_k})
\end{equation}

where

\begin{equation}
\mu_k^j = (\mu_{u_1}^j, \ldots, \mu_{u_{4^k}}^j)
\end{equation}

\begin{equation}
\sigma_k^j = (\sigma_{u_1}^j, \ldots, \sigma_{u_{4^k}}^j)
\end{equation}

and

\begin{equation}
\dfrac{\mu_k^j}{\sigma_k^j} = (\dfrac{\mu_{u_1}^j}{\sigma_{u_1}^j}, \dots, \dfrac{\mu_{u_{4^k}}^j}{\sigma_{u_{4^k}}^j}).
\end{equation}

\begin{figure}[ht]
\centering
\includegraphics[width=3.5in]{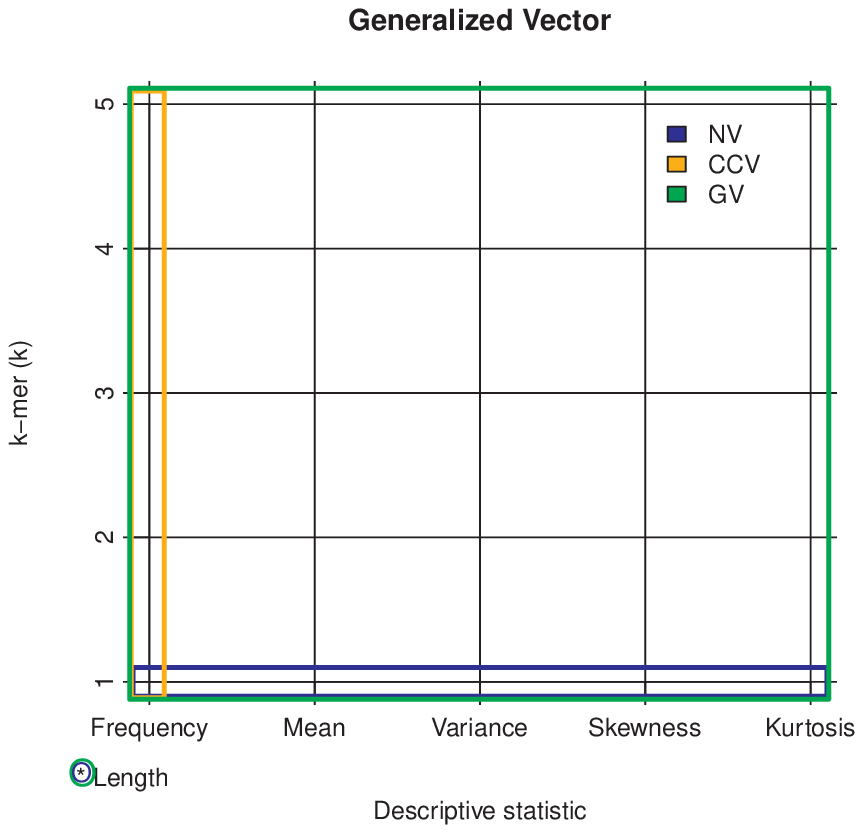}
\caption{The descriptive space of genome vectorizations.}
\label{f:gnv}
\end{figure}

Figure \ref{f:gnv} shows the approximate descriptive space occupied by the various vectorizations.  The complete composition vector uses the frequencies but ignores all additional position information and throws out length.  The natural vector uses counts and so length is described, in addition to mean, variance, and higher-order descriptive statistics that can be transformed to describe skewness and kurtosis.  The generalized vector uses the length in addition to the frequency, mean, variance, etc. of all k-mers.

\subsubsection{One-to-One}

In \cite{modeng} the authors show that there is a one-to-one correspondence between a genome and its natural vector.  The same is true for k-mers with $k = n$.  That is, for a genome of length $n$ and a k-mer vector with $k=n$, there is exactly one k-mer in the $4^k$ length vector that is non-zero.  The generalized vector maintains the one-to-one correspondance given
that one may fix $k \geq 1$ and let $j=\max\{n_{u_1}, \ldots, n_{u_{4^k}}\}$ which guarantees one-to-one correspondence.  In practice, we use $k \leq 5$ and $j \leq 4$.

\newpage

\section{Algorithm}
\label{s:analysis}

\subsection{Phylogenetic Classes}

Viruses are classified phylogenetically using two complementary systems.  The first system is known as \textit{Baltimore classification}~\citep{baltimorepaper}.  Baltimore classifications are defined by the genomic material of the virus (RNA/DNA), strandedness (single/double), the method of replication (reverse-transcribing), and sense (positive/negative).  This results in 7 mutually exclusive viral classes.

The \textbf{International Committee on Taxonomy of Viruses} (ICTV) provides the second method of classification~\citep{ictvpaper}.  The classifications are made by a sub-committee of the ICTV based on features of the virus (e.g. capsid shape, host, genome sequence, etc.)  These classifications are hierarchical.  The levels of the hierarchy, ordered from the broadest to the most specific, are \textit{order}, \textit{family}, \textit{sub-family}, \textit{genus}, \textit{species}.  Additionally, each family belongs to only one Baltimore class.  There are additional levels of the hierarchy, e.g. \textit{sub-genus}, but for the data used here only the Baltimore class, family, genus, and species are analyzed.

\subsection{Training and Testing}

Each dataset is split up randomly into a training set of 75\% of the data and a testing set of the remaining 25\%.  The same cross-validation folds (training) and testing sets are used for all of the vectorizations.

Since we perform cross-validation to determine optimal parameters, and because some of the labels are small in number, it is required that a class label have at least 3 samples; 1 sample for testing and 2 for training.  Classes with fewer than 3 samples are removed.  In practice, the viruses in these classes can be added back into the training set for the final model.  The procedure for determining if a virus belongs to a new class is discussed below.  We also require proportional distribution of the classes amongst the training and testing sets in addition to proportional distribution amongst the cross-validation splits.  We use 10-fold cross validation where possible, and smaller where it is not. 

\subsection{Data}

The two data sets used are the Reference Sequence data (RefSeq) published by the National Center for Biotechnology Information (NCBI) and the PASC data.  The RefSeq data consists of over 2000 viruses, but after removing viruses with multiple segments or without Baltimore classes, only 1881 viruses remain.

The PASC data consist of 51 families with 8862 viruses in total.  These data are used to predict species since that is the primary objective of the web tool.

\subsection{PASC}
\label{pascsec}

The PASC web tool uses a BLAST-based alignment method.
The precomputed similarity scores were downloaded, and are accessible, from the PASC website~\citep{pascweb}.  PASC matrices are not calculated for the RefSeq data and the method is ignored for that evaluation.

\subsection{$k$-Nearest Neighbors}

The restriction of PASC to similarity matrices resulted in the $k$-nearest neighbor algorithm being the most straightforward to implement.  The value of $k$ within the $k$-nearest neighbor algorithm is chosen by cross-validation.  

\subsection{Relevant Component Analysis}
\label{sec:RCA}

With regards to GV and CCV, the exponential growth of the vector size for larger values of $k$ within k-mers ensures that the smaller values of $k$ will be overwhelmed by the larger values of $k$; e.g. there are only 4 1-mers while there are 1024 5-mers.  For this reason we perform a version of relevant component analysis (RCA) to (1) improve classification accuracy and (2) because the transformations may provide valuable information for practitioners.

Where the standard RCA \citep{rca} takes the average of the absolute value of a component's correlation amongst all labels, we instead use cross-validation to:

(1) take the absolute value of the correlation to some power between 0 and 10 before taking the average and

(2) we enforce some sparsity by reducing to 0 some percentage of the smallest coefficients.  

\subsection{Partitions}

We perform the above analysis on each dataset 5 times using 5 randomly chosen testing and training set partitions to ensure the reliability of the results.
From the single-segment 2044 RefSeq viruses, 
1881 viruses are used for training (1413) and testing (468) in total.
For each partition of the PASC data there are 5559 training samples and 1758 testing samples.

\subsection{Cross-validation}

Cross-validation is used to tune the parameters of a model.  Typically, this is done by performing a grid search over a reasonable parameter space~\citep{10}.  In \cite{bergstra2012random} a randomized search is shown to be a more efficient method and is used here. 

\subsection{Predictions and Errors}

Within the PASC data evaluations, predicted class labels are recorded.  Viruses where the predicted class labels do not match the labels given in the NCBI or PaSC datasets are assumed to be errors.  While this is not always true due to the inherently messy nature of the data, the low error rates described below indicate that the overwhelming majority of the species labels are reliable.

\newpage

\section{Implementation}
\label{s:eval}

\subsection{Reference Sequence Results}

For Baltimore classifications, with results shown in Table \ref{baltpred}, GV performs the best and has an average misclassification rate of 2.9\% over the 5 partitions.  CCV, NV, and CV have average misclassification rates of 6.8\%, 8.2\%, and 11.8\% respectively.

Results for family classifications given the Baltimore class are shown in Table \ref{fampred}.  GV again performs the best and has an average misclassification rate of 5.5\% over the 7 Baltimore classes and 5 partitions compared to 8.9\%, 13.3\%, and 14.7\% misclassification rates for CCV, CV, and NV respectively.

Results for genus classifications given family labels are shown in Table \ref{genuspred}.  GV again performs the best, but this time it ties with CCV with an average misclassification rate over the 72 families and 5 partitions of 5.7\% compared to 8.4\% and 12.3\% misclassification rates for CV and NV respectively.

\subsection{PASC results}

The totals on the bottom of Table 4 
show that CCV and GV are both very competitive with PASC on this data hand-picked for PASC with error rates of 0.7\% and 0.8\%, respectively, compared to PASC's 0.6\%.  CV and NV on the other hand, struggle in many cases.  Additionally, the PASC webtool is not portable in the sense that it relies on NCBI resources and cannot be implemented on a PC.  The other four methods can be utilized on a PC easily.

One case where GV noticeably underperforms compared to PASC and CV is in the family Picornaviridae, with 9 errors total.  While this bears more investigation the error rate within that family remains below 1.2\%.  For CCV and GV, the error rates never exceed 4\% on any virus family, reaching their maximum in the Paramyxoviridae and Togaviridae families respectively.  PASC's error rate within families reaches its maximum in the Hepadnaviridae family with 6.67\%.



\newpage

\section{Discussion}
\label{sg0:concl}

We have generalized the class of genome statistics for sequences that comprise the vectorizations used for phylogenetic classification, thereby avoiding the troubles that accompany sequence alignment.  The performance of the GV is superior to the other vectorizations on Baltimore and family classifications.  On genus-level and species-level classifications GV performs as well as, or almost as well as, CCV and PASC.

The coefficients generated by the RCA methodology are simple and intuitive, but other methodologies may be more effective; e.g. principle component analysis~\citep{jolliffe2005principal}, neighborhood component analysis~\citep{goldberger2004neighbourhood}, or large-margin nearest neighbors~\citep{blitzer2005distance}.  PASC requires a two-step process that requires first identifying the appropriate virus family.  Additionally, PASC requires the use of high-performance computing that may not be available in low-resource environments.  The GV method described here requires less than a second to classify new viruses using existing models and less than a minute to generate entirely new models on a consumer laptop.

Future work could include the GV being extended to maximal length using the suffix-tree methods that have already been shown to be effective with CCA methods in phylogenetic classification~\citep{apostolico2010efficient}.  Additionally, the method described above should be considered a proof-of-concept.  The determination of new virus classes (and incorrect labels) can be handled in practice using techniques developed in the deep $k$-nearest neighbor literature~\citep{denceuxk}, one-class SVMs~\citep{chen2001one}, and cluster analysis~\citep{tibshirani2001estimating}.

Taking classification performance and computational performance features into account, the GV method provides a useful alternative to PASC for phylogenetic classification.  Given the many and varied applications of k-mers, this new class of genome statistics may prove to be additionally useful outside the field of phylogenetics.

\newpage

\section{Funding}
This work was supported by National Science Foundation [DMS-1120824], University of Illinois at Chicago, and Tsinghua University.

\newpage

\section{Acknowledgement}
Dr.~Yiming Bao and Dr.~Chenglong Yu provided helpful suggestions.

\newpage

\section{Author Disclosure Statement}

The corresponding author has previously contacted the editor, Michael Waterman, concerning collaboration.

\newpage

\bibliographystyle{jcb}
\bibliography{references2}

\newpage

\begin{table}[]
\begin{center}
\begin{tabular}{|l|rrrrrrr|r|}
  \hline
 & I & II & III & IV & V & VI & VII & Totals \\
  \hline
\# Train & 582 & 246 & 34 & 423 & 51 & 44 & 33 & 1413 \\
  \# Test & 194 & 82 & 11 & 140 & 16 & 14 & 11 & 468 \\
  \# Remv'd & 0 & 0 & 0 & 0 & 0 & 0 & 0 & 0 \\
  \# Total & 776 & 328 & 45 & 563 & 67 & 58 & 44 & 1881 \\
  \hline
  NV Errors & 4.8 & 7.0 & 4.4 & 10.4 & 3.2 & 6.6 & 1.8 & 38.2 \\
  CV Errors & 4.8 & 20.4 & 7.8 & 17.2 & 1.4 & 1.8 & 1.8 & 55.2 \\
  CCV Errors & 2.6 & 13.4 & 4.8 & 8.2 & 1.0 & 1.2 & 0.2 & 31.4 \\
  GV Errors & 1.6 & 5.8 & 2.4 & 1.8 & 0.6 & 0.8 & 0.4 & 13.4 \\
   \hline
\end{tabular}
\caption{Baltimore errors and samples averaged over 5 partitions.}
\label{baltpred}
\end{center}
\end{table}

\clearpage

\newpage

\begin{table}
\begin{center}
\begin{tabular}{|l|rrrrrrr|r|}
  \hline
 & I & II & III & IV & V & VI & VII & Totals \\ 
  \hline
\# Train & 558 & 238 & 28 & 400 & 48 &   0 & 33 & 1305 \\ 
  \# Test & 178 & 76 & 8 & 124 & 15 &   0 & 11 & 412 \\ 
  \# Remv'd & 40 & 14 & 9 & 39 & 4 & 58 & 0 & 164 \\ 
  \# Total & 776 & 328 & 45 & 563 & 67 & 58 & 44 & 1881 \\ 
  \hline
  NV Errors & 42.8 & 3.0 & 1.0 & 13.0 & 0.6 &   0.0 & 0.0 & 60.4 \\ 
  CV Errors & 27.4 & 3.2 & 1.4 & 21.6 & 1.2 &   0.0 & 0.0 & 54.8 \\ 
  CCV Errors & 23.6 & 2.6 & 1.0 & 8.8 & 0.6 &   0.0 & 0.0 & 36.6 \\ 
  GV Errors & 17.0 & 0.6 & 0.4 & 4.4 & 0.2 &   0.0 & 0.0 & 22.6 \\ 
   \hline
\end{tabular}
\caption{Family errors and samples given Baltimore class averaged over 5 partitions.}
\label{fampred}
\end{center}
\end{table}

\clearpage

\newpage

\begin{table}
\begin{center}
\begin{tabular}{|l|rrrrrrr|r|}
  \hline
 & I & II & III & IV & V & VI & VII & Totals \\ 
  \hline
\# Train & 252 & 221 & 16 & 330 & 32 & 43 & 23 & 917 \\ 
  \# Test & 77 & 67 & 4 & 101 & 10 & 10 & 7 & 276 \\ 
  \# Remv'd & 447 & 40 & 25 & 132 & 25 & 5 & 14 & 688 \\ 
  \# Total & 776 & 328 & 45 & 563 & 67 & 58 & 44 & 1881 \\ 
  \hline
  NV Errors & 10.6 & 2.4 & 1.4 & 10.4 & 2.4 & 2.4 & 1.2 & 30.8 \\ 
  CV Errors & 2.8 & 1.8 & 2.2 & 10.8 & 2.0 & 0.4 & 1.0 & 21.0 \\ 
  CCV Errors & 2.0 & 2.4 & 1.2 & 7.0 & 0.6 & 0.0 & 1.2 & 14.4 \\ 
  GV Errors & 3.0 & 1.6 & 1.6 & 6.0 & 1.2 & 0.2 & 0.8 & 14.4 \\ 
   \hline
\end{tabular}
\caption{Genus errors and samples given family class averaged over 5 partitions.}
\label{genuspred}
\end{center}
\end{table}

\clearpage

\newpage

\begin{longtable}{|r|llll|lllll|}
  \hline
 & \# Train & \# Test & \# Remv'd & \# Total & CV & CCV & NV & PASC & GV \\
   \hline
 \endhead

  \hline
 Adenoviridae & 70 & 22 & 31 & 123 & 0.0 & 0.0 & 0.2 & 0.0 & 0.0 \\
  Alloherpesviridae & 0 & 0 & 5 & 5 & 0.0 & 0.0 & 0.0 & 0.0 & 0.0 \\
  Alphaflexiviridae & 66 & 20 & 39 & 125 & 0.0 & 0.0 & 0.0 & 0.0 & 0.0 \\
  Anelloviridae & 151 & 49 & 38 & 238 & 0.0 & 0.0 & 0.4 & 0.0 & 0.0 \\
  Arteriviridae & 120 & 39 & 3 & 162 & 0.0 & 0.0 & 0.0 & 0.0 & 0.0 \\
  Astroviridae & 26 & 8 & 15 & 49 & 0.0 & 0.2 & 0.2 & 0.0 & 0.2 \\
  Avsunviroidae & 292 & 95 & 0 & 387 & 0.0 & 0.2 & 0.0 & 0.0 & 0.2 \\
  Baculoviridae & 8 & 3 & 52 & 63 & 0.0 & 0.0 & 0.6 & 0.0 & 0.0 \\
  Betaflexiviridae & 73 & 22 & 46 & 141 & 0.0 & 0.4 & 0.2 & 0.0 & 0.0 \\
  Caliciviridae & 227 & 73 & 10 & 310 & 0.8 & 1.4 & 1.2 & 1.2 & 0.8 \\
  Caulimoviridae & 33 & 10 & 52 & 95 & 0.0 & 0.0 & 0.2 & 0.0 & 0.0 \\
  Circoviridae & 272 & 88 & 18 & 378 & 0.0 & 0.0 & 0.4 & 0.0 & 0.0 \\
  Coronaviridae & 108 & 34 & 28 & 170 & 0.8 & 0.0 & 0.4 & 0.0 & 0.2 \\
  Dicistroviridae & 21 & 7 & 13 & 41 & 0.0 & 0.0 & 0.0 & 0.0 & 0.0 \\
  Endornaviridae & 0 & 0 & 11 & 11 & 0.0 & 0.0 & 0.0 & 0.0 & 0.0 \\
  Filoviridae & 20 & 6 & 3 & 29 & 0.0 & 0.0 & 0.0 & 0.0 & 0.0 \\
  Flaviviridae & 562 & 183 & 41 & 786 & 2.8 & 0.6 & 4.8 & 0.4 & 0.6 \\
  Geminiviridae & 505 & 154 & 220 & 879 & 3.0 & 3.8 & 14.0 & 4.0 & 3.4 \\
  Hepadnaviridae & 50 & 15 & 8 & 73 & 0.6 & 0.4 & 1.0 & 1.0 & 0.0 \\
  Herpesviridae & 8 & 2 & 55 & 65 & 0.0 & 0.0 & 0.0 & 0.0 & 0.0 \\
  Hypoviridae & 0 & 0 & 9 & 9 & 0.0 & 0.0 & 0.0 & 0.0 & 0.0 \\
  Iflavirus & 13 & 3 & 7 & 23 & 0.0 & 0.0 & 0.0 & 0.0 & 0.0 \\
  Inoviridae & 0 & 0 & 38 & 38 & 0.0 & 0.0 & 0.0 & 0.0 & 0.0 \\
  Iridoviridae & 6 & 2 & 10 & 18 & 0.0 & 0.0 & 0.0 & 0.0 & 0.0 \\
  Lentivirus & 699 & 230 & 10 & 939 & 4.4 & 1.2 & 6.0 & 0.8 & 1.6 \\
  Leviviridae & 23 & 6 & 3 & 32 & 0.4 & 0.0 & 0.2 & 0.0 & 0.0 \\
  Lipothrixviridae & 0 & 0 & 8 & 8 & 0.0 & 0.0 & 0.0 & 0.0 & 0.0 \\
  Luteoviridae & 73 & 22 & 19 & 114 & 0.0 & 0.4 & 0.0 & 0.0 & 0.4 \\
  Microviridae & 44 & 13 & 15 & 72 & 0.0 & 0.0 & 0.0 & 0.0 & 0.0 \\
  Nanoviridae\_CP & 25 & 8 & 6 & 39 & 0.0 & 0.0 & 0.0 & 0.0 & 0.0 \\
  Nanoviridae\_Rep & 0 & 0 & 48 & 48 & 0.0 & 0.0 & 0.0 & 0.0 & 0.0 \\
  Narnaviridae & 0 & 0 & 13 & 13 & 0.0 & 0.0 & 0.0 & 0.0 & 0.0 \\
  Papillomaviridae & 157 & 49 & 86 & 292 & 4.6 & 0.4 & 5.0 & 0.0 & 0.4 \\
  Paramyxoviridae & 168 & 51 & 17 & 236 & 2.4 & 1.8 & 1.6 & 2.0 & 1.2 \\
  Parvoviridae & 84 & 24 & 62 & 170 & 0.6 & 0.2 & 2.4 & 0.2 & 0.2 \\
  Picornaviridae & 491 & 155 & 39 & 685 & 4.8 & 0.2 & 4.8 & 0.0 & 1.8 \\
  Podoviridae & 7 & 2 & 113 & 122 & 0.0 & 0.0 & 0.0 & 0.0 & 0.2 \\
  Polyomaviridae & 109 & 34 & 28 & 171 & 0.0 & 0.0 & 0.4 & 0.0 & 0.4 \\
  Pospiviroidae & 491 & 155 & 8 & 654 & 1.4 & 1.2 & 3.0 & 0.6 & 1.2 \\
  Potyviridae & 209 & 66 & 59 & 334 & 1.0 & 0.0 & 0.4 & 0.0 & 0.0 \\
  Poxviridae & 8 & 3 & 30 & 41 & 0.0 & 0.0 & 0.0 & 0.0 & 0.0 \\
  Rhabdoviridae & 87 & 28 & 27 & 142 & 0.2 & 0.0 & 0.0 & 0.0 & 0.0 \\
  SecoviridaeRNA1 & 21 & 7 & 34 & 62 & 0.8 & 0.0 & 0.0 & 0.0 & 0.0 \\
  Sobemovirus & 28 & 9 & 13 & 50 & 0.0 & 0.0 & 0.0 & 0.0 & 0.0 \\
  Tectiviridae & 0 & 0 & 8 & 8 & 0.0 & 0.0 & 0.0 & 0.0 & 0.0 \\
  Tobamovirus & 78 & 23 & 22 & 123 & 0.0 & 0.4 & 0.2 & 0.0 & 0.4 \\
  Togaviridae & 92 & 26 & 13 & 131 & 1.4 & 0.2 & 2.0 & 0.4 & 1.0 \\
  Tombusviridae & 18 & 6 & 48 & 72 & 0.0 & 0.0 & 0.0 & 0.0 & 0.0 \\
  Totiviridae & 4 & 2 & 32 & 38 & 0.0 & 0.0 & 0.0 & 0.0 & 0.0 \\
  Tymoviridae & 5 & 2 & 29 & 36 & 0.0 & 0.0 & 0.0 & 0.0 & 0.0 \\
  Umbravirus & 7 & 2 & 3 & 12 & 0.0 & 0.0 & 0.0 & 0.0 & 0.0 \\
  \hline
  Totals & 5559 & 1758 & 1545 & 8862 & 30.0 & 13.0 & 49.6 & 10.6 & 14.2 \\
  \hline

\caption{Errors and Samples by Family Averaged Over 5 Partitions}
\end{longtable}
\label{tbl:pascresults}


\end{document}